# Boosting Objective Scores of a Speech Enhancement Model by MetricGAN Post-processing

*Szu-Wei Fu*[1,2,*], *Chien-Feng Liao*[2,*], *Tsun-An Hsieh*[2,*], *Kuo-Hsuan Hung*[2,*], *Syu-Siang Wang*[2], *Cheng Yu*[2], *Heng-Cheng Kuo*[2], *Ryandhimas E. Zezario*[1,2], *You-Jin Li*[2], *Shang-Yi Chuang*[2], *Yen-Ju Lu*[2], *Yu-Chen Lin*[1,2], *Yu Tsao*[2]

[1] Department of Computer Science and Information Engineering, National Taiwan University, Taipei, Taiwan
[2] Research Center for Information Technology Innovation, Academia Sinica, Taipei, Taiwan

{jasonfu,yu.tsao}@citi.sinica.edu.tw

## Abstract

The Transformer architecture has demonstrated a superior ability compared to recurrent neural networks in many different natural language processing applications. Therefore, our study applies a modified Transformer in a speech enhancement task. Specifically, positional encoding in the Transformer may not be necessary for speech enhancement, and hence, it is replaced by convolutional layers. To further improve the perceptual evaluation of the speech quality (PESQ) scores of enhanced speech, the $L_1$ pre-trained Transformer is fine-tuned using a MetricGAN framework. The proposed MetricGAN can be treated as a general post-processing module to further boost the objective scores of interest. The experiments were conducted using the data sets provided by the organizer of the Deep Noise Suppression (DNS) challenge. Experimental results demonstrated that the proposed system outperformed the challenge baseline, in both subjective and objective evaluations, with a large margin.

**Index Terms**: speech enhancement, PESQ, MetricGAN, DNS challenge

## 1. Introduction

Commercial speech-related applications, such as automatic speech recognition (ASR), hearing aids systems, and VoIP services rely heavily on clear sound provided by robust speech enhancement (SE) systems [1–4]. A SE system aims to reduce the background noise from noisy speech signals and further improve the quality and intelligibility of enhanced signals. Traditional approaches utilize the statistical attributes of speech signals for enhancement under many circumstances. However, these approaches require certain premises. For instance, a widely used denoise approach, Wiener filtering [5], performs well in many conditions, but the input must be guaranteed as a stationary process, which may not be fulfilled in a real-world situation.

Deep learning algorithms are known for their powerful capability of learning transformations. Learned features are usually more representative than handcrafted ones. In recent years, deep learning algorithms have been incorporated into SE tasks. Some approaches [6–10] operate SE on time-frequency acoustic features provided by short-time Fourier transform (STFT). Lu et al. [3] used a deep denoising autoencoder (DDAE) to estimate the enhanced speech. A main drawback of DDAE is that global time information is not considered because it merely depends on the frames nearby to predict an enhanced frame, and not on the entire sequence. To avoid this problem, some approaches operate SE in a sequential modeling manner. Weninger et al. [11] and Maas et al. [12] utilized recurrent neural networks (RNNs) for SE systems and further improved the robustness of ASR systems. RNNs, such as long short-term memory (LSTM) and gated recurrent unit, handle multiple gates and use hidden states to capture correlations within a sequence. However, it is difficult for RNNs to learn long-range dependencies between symbols because of sequential processing. In addition, the computation of gates is inefficient because of the time dependency between each other.

Therefore, the Transformer model has been proposed and has been demonstrated to achieve state-of-the-art results in various natural language processing tasks [13]. Specifically, to model long-range dependencies, the sequence relation between all-time steps is learned using the parallelly-computed attention mechanism [14]. Unlike RNNs, the Transformer processes an input sequence in parallel, thus significantly increasing training and inference efficiency. However, Kim et al. [15] found that the original Transformer did not demonstrate improvements in speech enhancement tasks. Hence, they proposed Gaussian-weighted self-attention and surpassed the LSTM-based model. In our study, we found that positional encoding in Transformer might not be necessary for SE, and hence, it was replaced by convolutional layers.

To further boost the objective scores of speech enhanced using the modified Transformer model, we applied it as a generator to the previously proposed MetricGAN [16]. Using some training techniques proposed by [17], the MetricGAN framework was used as a post-processing module. Specifically, the SE model (generator) was first pre-trained with a conventional loss function (e.g., $L_1$ or $L_2$ loss) until it converged, and the surrogate loss from the discriminator further guided the generator training to achieve a better solution. Because previous studies [16, 17] have already successfully applied BLSTM and convolutional BLSTM as the generator of MetricGAN, this framework can be treated as a general module to improve the performance of a trained deep SE model.

*Equal Contribution

## 2. Transformer Model for Causal Speech Enhancement

In our study, we developed the Transformer as the backbone architecture of our SE system and made some modifications to fit the denoising task. In addition, we adopted a causal setting to achieve the real-time processing requirement. The original Transformer consisted of encoder and decoder networks for sequence-to-sequence learning. First, for SE, the decoder part was omitted from our system because the input and the output sequence lengths were identical. Second, to inject some type of relative location information into the frames in the sequence, causal convolutional layers were utilized, instead of the original positional encoding. There are many choices of positional encodings, learned and fixed [18]. The learned ones require the input sequence to be of fixed length and are thus unable to adapt to sequence lengths that are longer than those encountered during training. Although the fixed ones may allow the model to adapt to variable sequence lengths, hand-crafted fixed features may not be rich enough for embedding. Hence, for both flexibility and model capacity, convolutional layers were chosen to capture location information. Finally, a future masking mechanism was applied to the scaled dot-product attention in the multi-head self-attention (MHSA) layer for causality, where the attention weights were set to zero for all future frames. More formally, three linear layers transformed the input argument of MHSA into queries $Q_h \in \mathbb{R}^{T \times d_k}$, keys $K_h \in \mathbb{R}^{T \times d_k}$, and values $V_h \in \mathbb{R}^{T \times d_k}$, where $T$, $h$ and $d_k$ denote the sequence length, head index, and feature dimension, respectively. The masked scaled dot-product is computed as

$$\text{Attention}(Q_h, K_h, V_h) = softmax(M + \frac{Q_h K_h^T}{\sqrt{d_k}})V_h \quad (1)$$

$M$ denotes the future masking, which is an upper triangular matrix, where the upper entries are set to negative infinity (excluding main diagonal), that is,

$$M = \begin{bmatrix} 0 & \cdots & -\infty \\ \vdots & \ddots & \vdots \\ 0 & \cdots & 0 \end{bmatrix} \quad (2)$$

In this way, future frames would not be considered because the upper entries of the attention weight became zero after the softmax function.

The remainder of the architecture is implemented as a standard Transformer shown in Figure 1, composed of N attention blocks. In each attention block, the first sub-layer is the masked MHSA, and the next is a feedforward network with two fully connected layers. Both sub-layers are followed by a residual connection to the input and layer normalization. Herein, the moments for layer normalization are computed only across the channel dimension, thus obeying the causal setting. Finally, the Transformer output is projected back to the frequency dimension using a fully connected layer with ReLU activation, and the $L_1$ loss is computed with clean speech.

### 2.1. Implementation

Speech waveforms were recorded at a 16 kHz sampling rate. A STFT with a hamming window size of 32 ms and a hop size of 16 ms was applied to transform the speech waveforms into 257-points spectral features. In the preliminary experiments, we found that compressing each coefficient to a tighter range produced better results; hence, the $log1p$ function ($log1p(x) = \log(1 + x)$) was adopted on the magnitude spectrogram. During testing, the enhanced spectral features were synthesized back to the waveform signals via the inverse STFT and an overlap-add procedure. The phases of the noisy signals were used for waveform generation. The Adam optimizer was used with a learning rate of 5e-5, and an early stopping was performed based on the validation set to prevent overfitting.

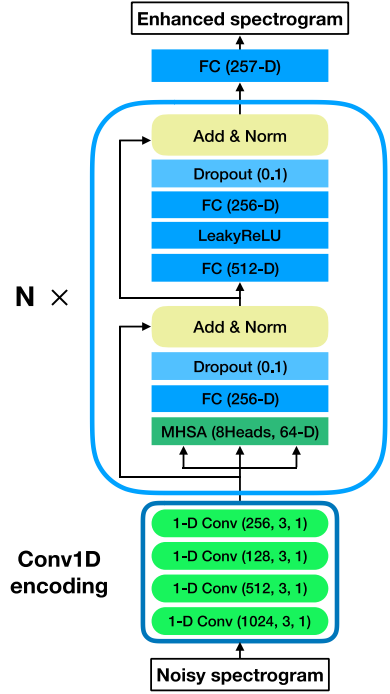

Figure 1: *Proposed Transformer architecture with 1-D convolutional encoding. 1-D Conv is in the format (output channels, kernel size, stride), and FC (output channels) denotes the fully connected layer. Add and Norm are the residual connections followed by layer normalization. Finally, each MHSA layer consists of 8 heads and 64 dimensions per head.*

## 3. Fine-tuning the Enhancement Model Using MetricGAN

Because the evaluation of this challenge is based on the ITU-T P.808, subjective evaluation of speech quality, a quality-related loss function may be a good choice to train the speech enhancement model. However, most of the quality metrics (e.g., PESQ) are too complicated to be directly applied as an objective function. Therefore, [19] employed a deep model (called Quality-Net [20]) to learn the behavior of PESQ function. Quality-Net is served as a surrogate to PESQ function to guide the enhancement model's learning. Although Quality-Net loss can further improve the PESQ score of enhanced speech, gradient provided by Quality-Net is only accurate for the first few learning iterations. In other words, Quality-Net is easily fooled (estimated quality scores increase but true scores decrease) [19].

To solve this problem, [16] proposed a learning framework such that Quality-Net and enhancement models are alternately updated. This method is called MetricGAN because its goal is to optimize black-box metric scores, and the architecture is similar to generative adversarial networks (GANs). Below, we briefly introduce the training of MetricGAN.

Let a function $Q'(I)$ represent the normalized evaluation metric (between 0 and 1) to be optimized, where $I$ denotes the input of the metric. For example, for PESQ and STOI, $I$ denotes a pair of enhanced speech, $G(x)$, that we want to evaluate and the corresponding clean speech, $y$. Therefore, to ensure that the discriminator network, ($D$), behaves similar to $Q'$, the objective function of $D$ is

$$L_{D(\text{MetricGAN})} = \mathbb{E}_{x,y}[(D(y,y) - 1)^2 + (D(G(x),y) - Q'(G(x),y))^2] \quad (3)$$

where $0 \leq Q'(G(x),y) \leq 1$.

The training of the generator network, ($G$), can completely rely on the adversarial loss

$$L_{G(\text{MetricGAN})} = \mathbb{E}_x[(D(G(x),y) - s)^2] \quad (4)$$

where $s$ denotes the desired assigned score. For example, to generate clean speech, we can simply assign $s$ to be 1.

Although the original MetricGAN is trained from scratch [16], Kawanaka *et al.* [17] proposed some training techniques to make MetricGAN a post-processing method of a trained speech enhancement model. After fine-tuning with the surrogate loss in MetricGAN, the metric scores of interest can be further improved. In our study, we applied some of the tricks to fine-tune the Transformer model (pre-trained with $L_1$ loss) to further boost the PESQ scores. The overall flow chart is shown in Figure 2.

## 4. Experiments

### 4.1. Dataset

The dataset used in this experiment was provided by the Deep Noise Suppression Challenge [21]. The default configuration was used to generate noisy-clean paired speech data. To reduce the training time, we randomly chose 10,000 training utterances to train our model. A synthetic test set without reverberation was selected as the validation set to evaluate the performance of different models. Subjective speech quality evaluation was based on a blind test set.

### 4.2. Model Structure

The pre-trained Transformer described in Section 2 is used as the SE model in our experiment. The parameters were first pre-trained with $L_1$-based signal approximation (SA) [11] loss. The discriminator (Quality-Net) was a CNN with four two-dimensional (2-D) convolutional layers with the number of filters and kernel size as follows: [15, (5, 5)], [25, (7, 7)], [40, (9, 9)], and [50, (11, 11)]. To handle the variable-length input, a 2-D global average pooling layer was added, so that the features were fixed with 50 dimensions (50 was the number of feature maps in the previous layer). Three fully connected layers were added subsequently, each with 50 and 10 LeakyReLU nodes and 1 linear node, respectively. To make Quality-Net a smooth function, we constrained it to be 1-Lipschitz continuous by spectral normalization [22].

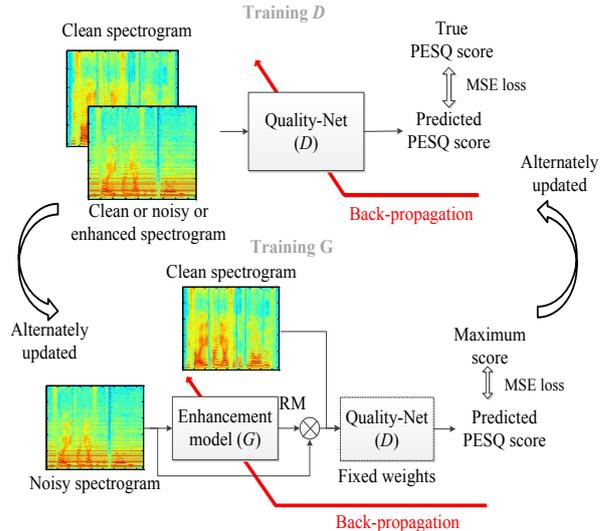

Figure 2: *Flow chart of MetricGAN training.*

Table 1: *Performance comparisons of different models on the synthetic test set without reverberation.*

|  | PESQ | STOI |
|---|---|---|
| Noisy | 2.454 | 0.915 |
| NSNet [23] | 2.692 | 0.906 |
| Transformer (PE) | 2.429 | 0.894 |
| Proposed Transformer | 2.966 | 0.932 |
| Transformer + MetricGAN | **3.104** | **0.946** |

### 4.3. Experimental Results of Objective Evaluation

To verify the proposed framework's effectiveness, the standard PESQ function was used to measure the speech quality, and the score ranged from −0.5 to 4.5. In addition, we presented STOI [24] for speech intelligibility evaluation, and the score ranged from 0 to 1. For both metrics, higher the score, better the quality. Table 1 presents the results of the average PESQ and STOI scores on the validation set for the Noise Suppression Net (NSNet) [8] baseline and the proposed method that fine-tunes the parameters of the Transformer model using the MetricGAN framework. As presented in Table 1, the significant performance drop in the Transformer with additive fixed positional encoding, denoted as Transformer (PE), echoes our hypothesis that the Transformer requires a better mechanism to inject location information. The performance of the proposed Transformer is much better than that of NSNet. When we pre-trained the Transformer model with $L_1$ loss and subsequently post-processed it using MetricGAN, we could further improve both the PESQ and STOI scores with a large margin. Note that, the computation load and model size remained unchanged because we only fine-tuned the parameters.

Figure 3 shows the fine-tuning process of the proposed MetricGAN. The PESQ score roughly converges to 2.97 when the $L_1$ loss is used to train the Transformer model. When the surrogate loss in MetricGAN is applied, the score can be further improved by 0.13.

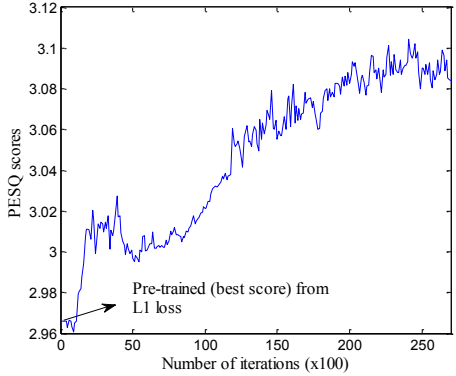

Figure 3: *PESQ scores of fine-tuning process using the MetricGAN framework.*

Table 2: *Computational complexity of the proposed model.*

|  | Number of parameters | Inference time (ms/frame) |
|---|---|---|
| Proposed Transformer | 5,953,920 | 0.256 |

### 4.4. Computational Complexity

In this section, we report the computational complexity in terms of the number of parameters and the time taken to infer a frame. As presented in Table 2, the number of weights in the proposed Transformer model is approximately 5.9 M, and it takes 0.256 ms to process a frame of 32 ms long (this is based on the average processing time of the whole blind test set) using an Intel Core i5 CPU quad core machine clocked at 2.4 GHz.

### 4.5. Spectrogram Comparison

Figure 4 shows the spectrograms of a clean utterance in the synthetic test set, the same utterance corrupted by traffic noise, enhanced speeches using the Transformer with $L_1$ loss, and fine-tuned using the proposed MetricGAN. From Figure 4(c), we observe that although the $L_1$ loss can guide the Transformer effectively removing the background noise, some noise still exists (as shown inside the blue dashed rectangle). Comparing figures 4(c) and 4(d), we find that the remaining noise can be further removed by the MetricGAN post-processing.

### 4.6. Subjective Evaluation

The organizer of the DNS challenge conducted a P.808 subjective evaluation of the submitted enhanced speech. 10 qualified judges rated each clip that resulted in a 95% confidence interval (CI) of approximately 0.02 on the overall mean opinion scores (MOS). The blind test set can be further split into noisy speech without reverberation (noreverb), noisy real recordings (realrec), and noisy reverberant speech (reverb) categories. Table 3 presents the MOS of noisy, NSNet baseline, and the proposed Transformer model fine-tuned using MetricGAN. From this table, we can observe that our proposed model can significantly outperform the baseline.

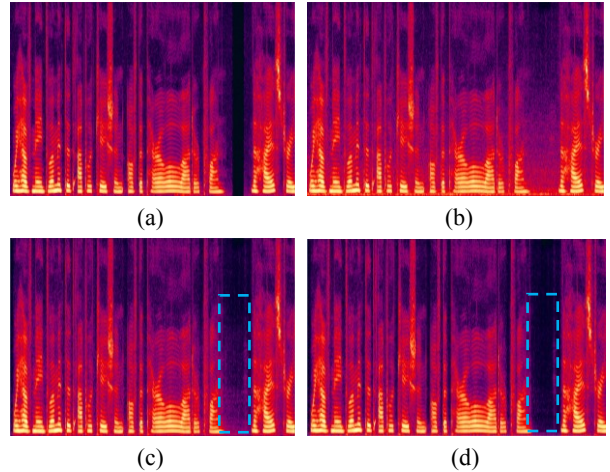

Figure 4: *Spectrograms of an utterance in the synthetic test set: (a) clean speech, (b) noisy speech (traffic noise), (c) enhanced speech using Transformer with $L_1$ loss (d) enhanced speech using Transformer + MetricGAN.*

Table 3: *MOS of the blind test set.*

|  | noreverb | realrec | reverb | Overall |
|---|---|---|---|---|
| Noisy | 3.32 | 2.97 | 2.78 | 3.01 |
| NSNet [23] | 3.49 | 3.00 | 2.64 | 3.03 |
| Transformer + MetricGAN | **3.63** | **3.18** | **2.83** | **3.21** |

## 5. Discussion

The proposed MetricGAN fine-tuning framework can be treated as a universal post-processing module for a speech enhancement model. For example, although we apply a Transformer as the generator (enhancement model) in our study, it works for other models such as BLSTM [16] and convolutional BLSTM [17] too. In addition to TF mask estimation, this method can improve the objective scores of the mapping-based enhancement model. However, to avoid generating additional artifacts (we observe that it may generate some high-frequency noise when we use PESQ or STOI function as $Q'$ because these two functions [24] ignore the signal difference in the high-frequency range.), we suggest that this post-processing is better applied to the mask estimation based deep model.

## 6. Conclusion

In our study, we applied a modified Transformer model as the generator of MetricGAN. To further boost the objective scores of interest, the Transformer model was first trained with the conventional $L_1$ loss, and then fine-tuned with the surrogate loss provided by the discriminator. Experimental results demonstrated that the proposed framework outperformed the challenge baseline, in terms of both objective scores and subjective evaluation. Using MetricGAN, we anticipate that the mismatch between human auditory perception and the loss used in training a speech enhancement model can be effectively reduced.